\documentclass[12pt]{article}

\usepackage{amsmath, amsthm, amssymb, enumerate,mathtools}
\usepackage{color}
\usepackage{datetime}
\usepackage{fancyhdr}
\usepackage{eufrak}
\chardef\coloryes=1 
\chardef\isitdraft=1 

\ifnum\isitdraft=1 
   \def\version{9} 
   \def\eqref#1{({\ref{#1}})}                

\definecolor{labelkey}{gray}{.3}

\definecolor{refkey}{rgb}{.3,0.3,0.3}

\else

  \def\startnewsection#1#2{\section{#1}\label{#2}\setcounter{equation}{0}}   
  \def\nnewpage{} 
\fi

 \usepackage{amsmath}
\usepackage{amsfonts}
\usepackage{amssymb}
\def\sg {{\sf g}}

\newtheorem{theorem}{Theorem}[section]
\newtheorem{lemma}{Lemma}[section]

\newtheorem{remark}{Remark}[section]
\newtheorem{definition}{Definition}[section]

\newtheorem{assu}{Assumption}[section]
\newtheorem{cor}{Corollary}[section]

\newcommand{\bgeqn}{\begin{eqnarray}}
\newcommand{\edeqn}{\end{eqnarray}}
\newcommand{\bgeq}{\begin{eqnarray*}}
\newcommand{\edeq}{\end{eqnarray*}}
\newcommand{\bec}{\begin{center}}
\newcommand{\enc}{\end{center}}
\newcommand{\beqn}{\begin{equation}}
\newcommand{\eeqn}{\end{equation}}
\newcommand{\belem}{\begin{lemma}}
\newcommand{\elem}{\end{lemma}}
\newcommand{\bethm}{\begin{theorem}} 
\newcommand{\ethm}{\end{theorem}}
\newcommand{\beitem}{\begin{itemize}}
\newcommand{\eitem}{\end{itemize}}
\newcommand{\beproof}{\begin{proof}}
\newcommand{\eproof}{\end{proof}}
\newcommand{\bedef}{\begin{definition}}
\newcommand{\edefn}{\end{definition}}
\newcommand{\beass}{\begin{assu}}
\newcommand{\eass}{\end{assu}}

\newcommand{\berem}{\begin{remark}}
\newcommand{\erem}{\end{remark}}
\newcommand{\pis}{\pi^\intercal_u \Sigma_u \pi_u}
\newcommand{\inc}{\intercal}

\newcommand{\Sg}{\Sigma}

\newcommand{\bbR}{\mathbb{R}}
\newcommand{\beas}{\begin{Assumption}}
\newcommand{\eas}{\end{Assumption}}

\newcommand{\Q}{{\cal Q}}

\newcommand{\F}{{\cal F}}

\newcommand{\B}{{\cal B}}

\newcommand{\bbP}{\mathbb{P}}
\newcommand{\bbN}{\mathbb{N}_0}

\newcommand{\be}{\begin{equation}}
\newcommand{\ee}{\end{equation}}
\newcommand{\rar}{\rightarrow}

\newcommand{\lan}{\langle }
\newcommand{\ran}{\rangle }

\def \intc {\intercal} 
\def \supmsg {\sup_{\gam \in \Gam}}
\def \supsg {\sup_{\gam \in \Gam}}

\def \Th{\Theta}
\def\dist{\mathop{\rm dist}}
\def\w{\omega}

\def\O{\Omega}

\def\eps{\varepsilon}

\def\supp {{\rm supp}}
\def\beal#1\eal{\begin{align}#1\end{align}}

\def \sg{{\sigma}}
\def \bext{\bbE^{\mathbb{P}_0}}

\def \gam {\gamma}
\def \Gam {\Gamma}

\newcommand{\argmax}{\operatornamewithlimits{arg\,max}}
\newcommand{\argmin}{\operatornamewithlimits{arg\,min}}

\def\bbe{{\Bbb{E}}} 

\def\bbP{{\mathbb{P}}}
\def\bbE{{\mathbb{E}}}
\def\hxt {\hat{X}_T^{\hat{\pi},\hat{\th}}}
\def\hxtp {\hat{X}_T^{\hat{\pi},\hat{\th}^*}}

\def\hpi{\hat{\pi}}

\def \rhot1{\rho_{t+1}}
\def \rhot{\rho_{t}}

\def \Sg {\Sigma}
\def \sg {\sigma}
\def \bbN{\mathbb{N}}

\def \O {\Omega}
\def \wo {\smallsetminus}
\def \eps {\epsilon}

\def \Diag {\mathrm{Diag}}
\def \pas {\bbP_0\mathrm{-a.s.}}

\def \bbS {\mathbb{S}}

\def \bqc {\bar{\Q}^c}

\def \treq {\triangleq}

\def \intzT {\int_{0}^T}

\def \intc{\intercal}

\def \th {\theta}
\def \Sth {S^{\th}}
\def \Qth {Q^{\th}}

\def \xpit {X^{\pi,\th}_T}

\def \ti {{t_i}}

\def \sgn {\sg_n}
\def \Piad {\Pi^{\mathrm{ad}}}

\def \piade    {\pi \in \Pi^{\mathrm{ad}}_{[t_{n-1},t_n)}}
\def \bbepn   {\bbE^{\bbP_0}_{t_n}}

\def \xpith   {X^{\pi,\th}_T}
\def \xpit   {X^{\pi,\th^*}_\tn}
\def \bepz    {\bbE^{\bbP_0}}
\def \inttnt  {\int^T_{t_n}}
\def \bextn   {\bepz_{t_n}}

\def \supin   {\sup_{\pi \in \Pi^{\mathrm{ad}}_{[t_n, T]}}\inf_{\th \in \Th_{\tnb}}}

\def \xpiT {X_T^{\pi,\th}}
\def \tn {{t_n}}
\def \supih {\sup_{\hat{\pi}\in \Pi^{\mathrm{ad}}_{[t_n,T]}}} 
\def \infih {\inf_{\hat{\pi}\in \Pi^{\mathrm{ad}}_{[t_n,T]}}}
\def \tib {[t_i,t_{i+1})}
\def \tnb {[t_n, T]}
\def \tne {[t_{n-1}, T]}
\def \tnet {[t_{n-1},t_n)}
\def \piadti {\Pi^{\mathrm{ad}}_{[t_i,t_{i+1})}}
\def \piadtn {\Pi^{\mathrm{ad}}_{[t_n, T]}}
\def \piadtne {\Pi^{\mathrm{ad}}_{[t_{n-1}, T]}}
\def \piadtnet {\Pi^{\mathrm{ad}}_{[t_{n-1}, t_n)}}
\def \Thtnb {\Theta_{[t_n, T]}}
\def \pisb   {\pi^\intercal_u \Sigma^*_{\tn} \pi_u}

 \renewcommand{\theequation}{\thesection.\arabic{equation}}

\begin{document}
\def\ques{{\cor \underline{??????}\cob}}
\def\nto#1{{\coC \footnote{\em \coC #1}}}
\def\fractext#1#2{{#1}/{#2}}
\def\fracsm#1#2{{\textstyle{\frac{#1}{#2}}}}   
\def\nnonumber{}


\def\cor{{}}
\def\cog{{}}
\def\cob{{}}
\def\coe{{}}
\def\coA{{}}
\def\coB{{}}
\def\coC{{}}
\def\coD{{}}
\def\coE{{}}
\def\coF{{}}
\ifnum\coloryes=1

  \definecolor{coloraaaa}{rgb}{0.1,0.2,0.8}
  \definecolor{colorbbbb}{rgb}{0.1,0.7,0.1}
  \definecolor{colorcccc}{rgb}{0.8,0.3,0.9}
  \definecolor{colordddd}{rgb}{0.0,.5,0.0}
  \definecolor{coloreeee}{rgb}{0.8,0.3,0.9}
  \definecolor{colorffff}{rgb}{0.8,0.3,0.9}
  \definecolor{colorgggg}{rgb}{0.5,0.0,0.4}

 \def\cog{\color{colordddd}}
 \def\cob{\color{black}}
 \def\cor{\color{red}}
 \def\coe{\color{colorgggg}}

 \def\coA{\color{coloraaaa}}
 \def\coB{\color{colorbbbb}}
 \def\coC{\color{colorcccc}}
 \def\coD{\color{colordddd}}
 \def\coE{\color{coloreeee}}
 \def\coF{\color{colorffff}}
 \def\coG{\color{colorgggg}}

\fi
\ifnum\isitdraft=1
   \chardef\coloryes=1 
   \baselineskip=17pt
\pagestyle{myheadings}
\reversemarginpar

\def\const{\mathop{\rm const}\nolimits}  
\def\diam{\mathop{\rm diam}\nolimits}    

 \def\llabel#1{\label{#1}{\ \mbox{\rm\color{red} {#1}\color{black}}}}

\def\rref#1{{\ref{#1}{\rm \tiny \fbox{\tiny #1}}}}
\def\theequation{\fbox{\bf \thesection.\arabic{equation}}}
\def\ccite#1{{\cite{#1}{\rm \tiny ({#1})}}}

\def\startnewsection#1#2{\newpage\cog \section{#1}\cob\label{#2}

\setcounter{equation}{0}
\pagestyle{fancy}

\lhead{\cob Section~\ref{#2}, #1 }
\cfoot{}
\rfoot{\thepage}
\lfoot{\cob{\today,~\currenttime}~(c75-iklt2, Version~\fbox{\version})}}
\chead{}
\rhead{\thepage}
\def\nnewpage{\newpage}

\newcounter{startcurrpage}
\newcounter{currpage}

\def\llll#1{{\rm\tiny\fbox{#1}}}
   \def\blackdot{{\color{red}{\hskip-.0truecm\rule[-1mm]{4mm}{4mm}\hskip.2truecm}}\hskip-.3truecm}
   \def\bdot{{\coC {\hskip-.0truecm\rule[-1mm]{4mm}{4mm}\hskip.2truecm}}\hskip-.3truecm}
   \def\purpledot{{\coA{\rule[0mm]{4mm}{4mm}}\cob}}
   \def\pdot{\purpledot}
\else
   \baselineskip=15pt
   \def\blackdot{{\rule[-3mm]{8mm}{8mm}}}
   \def\purpledot{{\rule[-3mm]{8mm}{8mm}}}
   \def\pdot{}
\fi

\def\nts#1{{\hbox{\bf ~#1~}}} 
\def\nts#1{{\cor\hbox{\bf ~#1~}}} 
\def\ntsf#1{\footnote{\hbox{\bf ~#1~}}} 
\def\ntsf#1{\footnote{\cor\hbox{\bf ~#1~}}} 
\def\bigline#1{~\\\hskip2truecm~~~~{#1}{#1}{#1}{#1}{#1}{#1}{#1}{#1}{#1}{#1}{#1}{#1}{#1}{#1}{#1}{#1}{#1}{#1}{#1}{#1}{#1}\\}
\def\biglineb{\bigline{$\downarrow\,$ $\downarrow\,$}}
\def\biglinem{\bigline{---}}
\def\biglinee{\bigline{$\uparrow\,$ $\uparrow\,$}}

\def\w{\omega}
\def\tilde{\widetilde}

\newtheorem{Theorem}{Theorem}[section]
\newtheorem{Corollary}[Theorem]{Corollary}
\newtheorem{Proposition}[Theorem]{Proposition}
\newtheorem{Lemma}[Theorem]{Lemma}
\newtheorem{Remark}[Theorem]{Remark}
\newtheorem{Example}[Theorem]{Example}
\newtheorem{Assumption}[Theorem]{Assumption}
\newtheorem{Claim}[Theorem]{Claim}
\newtheorem{Question}[Theorem]{Question}
\def\theequation{\thesection.\arabic{equation}}
\def\endproof{\hfill$\Box$\\}
\def\square{\hfill$\Box$\\}
\def\comma{ {\rm ,\qquad{}} }            
\def\commaone{ {\rm ,\qquad{}} }         
\def\dist{\mathop{\rm dist}\nolimits}    
\def\sgn{\mathop{\rm sgn\,}\nolimits}    
\def\Tr{\mathop{\rm Tr}\nolimits}    
\def\div{\mathop{\rm div}\nolimits}    
\def\supp{\mathop{\rm supp}\nolimits}    
\def\divtwo{\mathop{{\rm div}_2\,}\nolimits}    
\def\re{\mathop{\rm {\mathbb R}e}\nolimits}    
\def\div{\mathop{\rm{Lip}}\nolimits}   
\def\indeq{\qquad{}}                     
\def\period{.}                           
\def\semicolon{\,;} 
\mathtoolsset{showonlyrefs=true}


\title{Robust Utility Maximization with Drift and Volatility Uncertainty}
\author{Kerem U\u{g}urlu}
\maketitle

\date{}

\begin{center}
\end{center}

\medskip

\indent Department of  Mathematics, Nazarbayev University.\\
\indent e-mail:kerem.ugurlu@nu.edu.kz

\begin{abstract}
We give explicit solutions for utility maximization of terminal wealth problem $u(X_T)$ in the presence of Knightian uncertainty in continuous time $[0,T]$ in a complete market. We assume there is uncertainty on both drift and volatility of the underlying stocks, which induce nonequivalent measures on canonical space of continuous paths $\O$. We take that the uncertainty set resides in compact sets that are time dependent. In this framework, we solve the robust optimization problem with logarithmic, power and exponential utility functions, explicitly. 
\end{abstract}

\noindent\thanks{\em Mathematics Subject Classification\/}:
91B16;93E20\\
\noindent\thanks{\em Keywords:\/}
Knightian uncertainty; Mathematical finance; Optimal control

\section{Introduction}
Starting with the pioneering works of \cite{RO,B,AS0, RM,BSM}, the underlying risky assets are modelled as Markovian diffusions, where there exists a fixed underlying reference probability measure $\bbP$ that is retrieved from historical data of the price movements. However, it is mostly agreed that it is impossible to precisely identify $\bbP$. Hence, as a result, \textit{model ambiguity}, also called \textit{Knightian uncertainty}, in utility maximization is inevitably taken into consideration. Namely, the investor is diffident about the odds, and takes a robust approach to the utility maximization problem, where she minimizes over the priors, corresponding to different scenarios, and then maximizes over the investment strategies. 

The literature on robust utility maximization in mathematical finance, (see e.g. \cite{DW,MQ,AS,LSW, CE, MMDR,MMRU,AS,FS,R,RS,ES,GS,LS,AA,ADEH,CF,HT,GKM,GMM,KR} among others), mostly assumes that the set of priors is dominated by a reference measure $\bbP$. Hence, it presumes a setting where volatility of risky assets  are perfectly known, but drifts are uncertain. Namely, these approaches assume the equivalence of priors. In particular, they assume the equivalence of probability measures $P$ with a dominating reference prior $\mathbb{P}$.

A more general direction is the case, where the uncertainty on both mean and volatility is taken into consideration. Here, the set of priors are nondominated, and there exists no dominating reference prior $\mathbb{P}$. This approach started with the seminal works of \cite{ALP,L} in option pricing framework. In a more recent work, \cite{NZ} studied robust optimal stopping using nondominated measures, and its applications to subhedging of American options under volatility uncertainty. Regarding utility maximization, \cite{HS} studied the case, where uncertainty in the volatility is due to an unobservable factor. \cite{NN2} works in a jump-diffusion context, with ambiguity on drift, volatility and jump intensity. \cite{DK} establishes a minimax result and the existence of a worst-case measure in a setup where prices have continuous paths and the utility function is bounded. \cite{LR} works in a diffusion context, where uncertainty is modelled by allowing drift and volatility to vary in two constant order intervals. Here, the optimization using power utility of the from $U(x) = x^\gamma$ for $0 < \gamma <1$ is performed via a robust control (G-Brownian motion) technique, which requires the uncertain volatility matrix is diagonal. We refer the reader to \cite{P} for a detailed exposure on G-Brownian motion and its applications. \cite{MP} studies the utility maximization problem with power utility, where there is an ellipsoidal uncertainty for drift and volatility uncertainty that reside in a fixed compact set. \cite{MN} works in a continuous time setting, where the stock prices are allowed to be general discontinuous semi-martingales, and strategies are required to be compact and studies power utility and give semi-explicit solutions. \cite{MPZ} studies robust utility maximization in an incomplete market, where there exists a fixed compact uncertainty set for volatility and drift. They prove the existence of optimal strategies with power and utility functions using backward stochastic differential equations theory. \cite{FK} studies a general robust utility maximization problem, where it proposes to model a way to model drift and volatility. \cite{IP} studies the mean variance optimization in a diffusion setting, where it is assumed that the drift of the stock is known with certainty, whereas the volatility is assumed to be in some compact set. \cite{DB} shows the existence of optimal strategy in the robust exponential utility maximization problem in discrete time. 

On the other hand, we are studying a utility maximization problem in finite continuous time horizon in a diffusion setting, where there is time-dependent uncertainty on both drift and volatility residing in a compact set. Contrary to the usual stream that the compact set containing the differential characteristics is fixed throughout $[0,T]$, we assume that the set of priors is time dependent. There can be at least two arguments to support this construction. First, in an intraday movement of a stock, it is not reasonable to assume that drift and volatility uncertainty reside in a fixed compact set througout $[0,T]$. Second, with time drift and volatility of the stock can be \textit{learned} (see e.g. \cite{CK}) and hence the corresponding compact sets might change, as time proceeds. This more general approach entails additional technical problems. In particular, depending on the confidence set, the optimal value function might not be $C^{1,2}$, hence the classical Hamilton-Jacobi-Bellman-Ishii (HJBI) or the martingale optimality principle approach can not be used at the first place (see e.g. Theorem 1.1 \cite{RO}) and it requires a more careful analysis to overcome this hurdle.

The rest of the paper is as follows. In Section 2, we describe the model dynamics of the problem and state our general main problem and propose the solution methodology. In Section 3, we solve our utility maximization problem explicitly using logarithmic, power and exponential utility functions. In Section 4, we discuss our results and conclude the paper. 

\section{Model Dynamics and Investor's Value Function}

\subsection{Framework for Model Uncertainty and Model Dynamics}
We fix the dimension $d \in \bbN$ and time horizon $T \in (0,\infty)$. We let $\O = C_0([0,T])$ be the space of continuous paths $\w = (\w_t)_{0 \leq t\leq T}$ starting at $0 \in \bbR^d$. We define the coordinate functional for $\w \in \O$ as $W_t(\w) := \w_t$ and take the corresponding Borel $\sg$-algebra by $\F_t := \sg(W_s(\w):0 \leq s \leq t)$. We denote $\bbP_0$ as the Wiener measure on $\O$ such that $W_t$ is the $(\O,\F_t)$  Wiener process and take $\bbP_0$ as the \textit{reference} measure. We consider a market consisting of $d$ risky assets $S^{\th}_t= (S^{\th,1}_t,\ldots,S^{\th,d}_t)$ and one riskless asset $R_t$. We assume $S^\th_t$ and $R_t$ satisfy the following dynamics
\beal
dR_t &= rR_tdt\\
S_0 &= s_0\\
\label{eqn21500}
dS^\th_t &= \Diag(S^\th_t)(\mu_t dt  + \sg_t dW_t),\;\pas
\eal
Here $\Diag(S^\th_t)$ is a $d \times d$ diagonal matrix with $(S^{\th,1}_t,\ldots,S^{\th,d}_t)$ its diagonal entries. We take that $\mu_t$ is a progressively measurable $\bbR^d$-valued mapping, whereas $\sg_t$ is $d \times d$ matrix valued and progressively measurable. We further denote by $\Sg_t \treq \sg_t\sg^\intc_t$ the covarianc matrix of $d$ stocks.

\beas
\label{ass21}
We assume $0 \leq \lVert \mu_t \rVert \leq C^\mu_{t_i}$ $\pas$ and $0 < c^{\Sg,\min}_{t_i} \leq \lVert \Sg_t \rVert \leq C^{\Sg,\max}_{t_i}$ for $t_i \leq t < t_{i+1}$, $i = 0, \ldots,n-1$ and $0 \leq \lVert \mu_t \rVert \leq C_{\tn}$ $\pas$ for $\tn \leq t \leq T$, where $0= t_0< t_1 <\ldots < t_n < t_{n+1} = T$. We denote by $\Theta_{\tib} \subset \bbR^d \times \bbS^d_{+}$ the compact set containing the differential characteristics $\th_t \treq (\mu_t,\sg_t)$ for $\ti \leq t < t_{i+1}$ and $\Th_{\tnb} \subset \bbR^d \times \bbS^d_{+}$ for $t_n \leq t \leq T$. We further assume that there exists a strong solution the Equation \eqref{eqn21500} for any given $(\th_t)_{0 \leq t \leq T}$ on $(\O, \F_T, \bbP_0)$. Namely, denoting $C_0[0,T] = \O$ and $S^{\th}$ being as in Equation \eqref{eqn21500}, we take that there exists an $\F_T$ measurable mapping $G:\O \rar \O$ such that $S^\th(\cdot) \equiv G(x_{t_i},W(\cdot))$ solves Equation \eqref{eqn21500} on $(\O,\F_T, \bbP_0)$, as in Definition 10.9 in \cite{RW}.
\eas

\subsection{Alternative Models}
Note that for different $(\th_t)_{0 \leq t \leq T} \in (\Th_t)_{0 \leq t \leq T}$, different probability measures are induced on $\O$, which is defined as
\beal
\label{eqn270}
\Qth &\treq \bbP_0 \circ (\log(\Sth))^{-1},
\eal
where $S^\th$ has the differential characteristics as in Equation \eqref{eqn21500} with $(\th_t)_{0 \leq t \leq T} = (\mu_t,\sg_t)_{0 \leq t \leq T}$. 
Further, different $\sg_1,\sg_2$ induce nonequivalent probability measures. Indeed, for $\th_1 =(\mu_1,\sg_1)$ and $\th_2 = (\mu_2,\sg_2)$, where $(\mu_i,\sg_i)_{i=1,2}$ are constants in $\bbR^n$ and $\bbS^n_+$, respectively, we have
\beal
Q^{\th_1}(\lan \log(S^{\th_1}) \ran &= \sg_1\sg_1^\inc) = 1\\
Q^{\th_2}( \lan \log(S^{\th_2}) \ran &= \sg_2\sg_2^\inc) = 1.
\eal
Here $\lan \cdot \ran$ stands for the quadratic variation of $\log(S^\th)$. However, the dynamics of differential characteristics  are given with respect to $\bbP_0$, in particular, \textit{we look through the lenses of the Wiener measure} $\bbP_0$. 
This is possible, since we consider only strong solutions in Equation \eqref{eqn21500}.  

\subsection{Financial Scenario}
We consider the problem of an agent investing in $d$ risky assets $S_t$ and one riskless asset $R_t$. For a given initial endowment $x_0 > 0$, the investor trades in a self financing way. We denote $\hat{\pi}_t$ as an $n$-dimensional progressively measurable stochastic process, which stands for the total amount of money invested in $d$ risky assets $S_t$ at time $t$, $0 \leq t \leq T$. Then, we have for $X_0 = x_0 > 0$
\begin{align}
\label{eqnxexp}
d\hat{X}^{\hat{\pi},\th}_t &= \hat{\pi}^\inc_t S_t^{-1}\cdot dS_t + (\hat{X}^{\hat{\pi},\th}_t - {\hat{\pi}}^\inc_t \mathbf{1})rdt, \\
d\hat{X}^{\hat{\pi},\th}_t &= {\hat{\pi}}^\inc_t (\mu_t dt + \sigma_t \cdot dW_t) + (\hat{X}^{\hat{\pi},\th}_t - {\hat{\pi}}^\inc_t\mathbf{1})r dt\;\pas
\end{align}
We further represent the amount of money invested in $d$ risky assets as a fraction of current wealth via $\hat{\pi}_t =  \hat{X}^{\pi,\th}_t \pi_t$ for $0 \leq t \leq T$, where $\pi_t$ stands for the corresponding fraction at time $t$ and take the discounted wealth $X^{\pi,\th}_t = e^{-rt}\hat{X}^{\pi,\th}_t \pi_t$.  
Hence, for $X_0 = x_0$, the dynamics of wealth in this setting are given by 
\begin{align}
\label{eqn15}
d X_t^{\pi,\th} &=  X_t^{\pi,\th} \pi^\inc_t( (\mu_t-r\mathbf{1}) dt +  \sigma_t d W_t)\\
X_t^{\pi,\th} &= x_0\exp \int_0^t \pi^\inc_u (\mu_u-r\mathbf{1}) - \frac{1}{2} \pis du + \int_0^t \pi_u^\inc \sigma_u d W_u\;\pas,
\end{align}
where $\mathbf{1}$ stands for $d$ dimensional vector $(1,\dots,1)$. We further denote $X^{\pi,\th}$ as the wealth process with dynamics $(\th_t)_{0\leq t \leq T} = (\mu_t,\sg_t)_{0 \leq t \leq T}$ as in Equation \eqref{eqn15}. Here, $\Pi^{\mathrm{ad}}_{[0,T]}$ stands for the admissible portfolios on $[0,T]$ that are defined as follows.

\bedef
\label{def21}
Let $\{\pi_u\}_{0\leq u \leq T}$ denote the $\B([0,T])\otimes \F_{T}$ progressively measurable process representing the cash-value allocated in $d$ risky assets. We call $ (\pi_u)_{\{0 \leq u \leq T\}}$ admissible and denote it by $\pi \in \Pi^{\mathrm{ad}}_{[0,T]}$, if it satisfies
\beqn
X_t^\pi > 0,\; 0 \leq t \leq T,\;\pas 
\eeqn
Analogously, we denote by $\piadti$ for $0 =t_0 < t_1 < t_2 < \ldots < t_{i+1} \leq t_n  < T$ and $\piadtn$ the admissible cash values, if it satisfies
\beal
&X^\pi_t > 0,\; t_i \leq t < t_{i+1}, \\
&X^\pi_t > 0,\; t_n \leq t \leq T,
\eal 
respectively. 

\edefn

\subsection{Investor's Problem}
The investor utilizes the classical Merton problem, but she is also diffident about the underlying dynamics of the stocks both in terms of drift  $\mu_t$ and covariance matrix $\Sg_t$. She assumes that $\th_t \treq (\mu_t,\sg_t)$ is in some compact set $(\Th_t)_{0\leq t\leq T}$ standing for the priors on the underlying dynamics. The investor reevaluates its priors $(\Th_t)_{0 \leq t \leq T}$ on some prespecified times $0= t_0 < t_1 <\ldots < t_n < t_{n+1} = T$. At time $t_n$, we write the optimization problem of the investor for $X^\pi_{\tn}  = x_\tn$ as
\beqn 
\label{eqnval}
V(\tn,x_\tn) \treq \supin  \bextn \big[ u(X_T^{\pi,\th}) \big],
\eeqn 
where $\bextn[\cdot] \treq \bepz[ \cdot |\F_\tn]$. Hence, at time $t_0 = 0$ the optimization problem reads backwardly as 
\beal
V(t_0,x_0) &\treq \sup_{\pi \in \Pi^\mathrm{ad}_{[0,T]}} \inf_{\th \in \Th_{[0,T]}} \bepz[u(X_T^{\pi,\th})]\\
&= \sup_{\pi \in \Pi^\mathrm{ad}_{[t_0,t_1)}} \inf_{\th \in \Th_{[t_0,t_1)}}\bepz\big[\sup_{\pi \in \Pi^\mathrm{ad}_{[t_1,t_2)}} \inf_{\th \in \Th_{[t_1,t_2)}}\bepz_{t_1}\big[\\
&\indeq \ldots \sup_{\pi \in \Pi^\mathrm{ad}_{[t_n,T]}} \inf_{\th \in \Th_{[t_n,T]}}\bepz_{t_n}\big[u(X_T^{\pi,\th})\big]\ldots\big]
\eal
We continue with the following variant of so called Martingale Optimality Principle (see also Theorem 1.1 of \cite{RO}). 
\bethm {\label{MOP}} Suppose that 
\begin{enumerate}
\item[(A1)]
there exists a function $v:[t_n,T] \times \bbR^{+} \rar \bbR$, that is $C^{1,2}([\tn,T] \times \bbR^+)$ with $v(T,\cdot) = u(\cdot)$,
\item[(A2)] for any $\pi \in \Pi^{\mathrm{ad}}_{\tnb}$, there exists an optimal solution $\th \in \Th_{\tnb}$ of 
\beqn 
\inf_{\th \in \Th_{\tnb}}\bextn \big[ u(\xpith) \big]
\eeqn
such that 
\beqn 
\label{eqn2400}
Y_{t_n}^{\pi} \treq v(t_n,X^{\pi,\th}_{t_n})
\eeqn 
satisfies for $\tn\leq s \leq t\leq T$ 
\beqn
\bbe_s[Y_{t}^{\pi}] \leq Y^{\pi}_s, \pas.
\eeqn
\item[(A3)]
there exists some $\bar{\pi} \in \Pi^{\mathrm{ad}}_{\tnb}$ such that for $\tn\leq s \leq t\leq T$ 
\beqn
\bbe_s[Y_{t}^{\bar{\pi}}] = Y^{\bar{\pi}}_s, \pas
\eeqn
\end{enumerate}
Then, $\bar{\pi} \in \Pi^{\mathrm{ad}}_{\tnb}$ is optimal for the problem 
\beal 
\label{eqn2500}
V(t_n,x_\tn) &= \supin \bextn \big[ u(\xpiT)\big] \\
&= \bextn \big[ u(X^{\bar{\pi}}_T) \big]
\eal 
\ethm

\beproof
By (A1) and (A2), we have
\beal 
\bextn\big[Y^{\pi}_T\big] &= \bextn\big[ u(\xpiT)\big] \leq Y^{\pi}_{t_n},\; \pas\\
Y^{\pi}_{t_n} &= v(t_n, x) 
\eal
Taking supremum over $\Piad_{\tnb}$, we get 
\beqn 
V(t_n,x_\tn) = \sup_{\piadtn} \bextn[u(\xpiT)] \leq v(t_n,x_\tn). 
\eeqn 
By (A3), for some $\bar{\pi} \in \Pi^{\mathrm{ad}}_{\tnb}$, we have $\bbe_s[Y^{\bar{\pi}}_t] = Y^{\bar{\pi}}_s\;\pas$ for $t_n \leq s \leq t \leq T$. Then, 
\beal 
V(t_n,x_\tn) &= \bextn[Y^{\bar{\pi}}_{T}]\\
&= Y^{\bar{\pi}}_{t_n}\\
&= v(t_n,x_\tn).
\eal
Hence, we conclude the proof. 
\eproof
Applying Ito lemma for $t \geq \tn$ to $Y_t^{\pi}$ in \eqref{eqn2400}, we have by \eqref{eqn15}
\beal 
&dY^{\pi}_t = \bigg(v_t + v_x X^\pi_t \big( \pi_t^{\intc}(\mu_t - r\mathbf{1})\big) +(X^\pi_t)^2\frac{1}{2}v_{xx}\pi_t^{\intc}\Sg_t\pi_t \bigg)dt\\
&\indeq\indeq + X^\pi_t v_x\pi^{\intc}_t\sg dW_t,\; \ \pas
\eal 
By Theorem \ref{MOP}, for $x > 0$, \eqref{eqnval} satisfies the following HJBI PDE: 
\beal
\label{eqn270}
&\supin \bigg[ v_t + v_xx \pi_t^{\intc}(\mu_t -r\mathbf{1}) + \frac{1}{2}x^2v_{xx}\pi_t^{\intc}\Sg_t\pi_t \bigg] \\
& = v_t + x\supin \bigg[ v_x\pi_t^{\intc}(\mu_t -r\mathbf{1}) + x\frac{1}{2}v_{xx}\pi_t^{\intc}\Sg_t\pi_t \bigg] 
\\&= 0
\eal 
Similarly, for $X^{\pi,\th}_t$ in \eqref{eqnxexp}, we have 
\beal
\label{eqn27123}
&\supin \bigg[ v_t + v_x \pi_t^{\intc}(\mu_t -r\mathbf{1}) + \frac{1}{2}v_{xx}\pi_t^{\intc}\Sg_t\pi_t \bigg] \\
& = v_t + \supin \bigg[ v_x\pi_t^{\intc}(\mu_t -r\mathbf{1}) + \frac{1}{2}v_{xx}\pi_t^{\intc}\Sg_t\pi_t \bigg] 
\\&= 0
\eal 
\belem
The value function $V(t_n, x_\tn)$ as defined in Equation \eqref{eqn2500} is increasing and concave in $x_{t_n}$. 
\elem 
\beproof
Recall that by assumption, the utility function $u(\cdot)$ is increasing and concave and for \eqref{eqn15}
\beqn 
X_T^{\pi,\th} = x_{t_n}\exp \bigg[ \inttnt (\pi^\inc_u (\mu_u - r\mathbf{1}) - \frac{1}{2} \pis )du + \int_{t_n}^T \pi_u^\inc \sigma_u d W_u \bigg]\;,\pas
\eeqn 
In particular, for any $x^1_{t_n} \leq x^2_{t_n}$ with fixed $\Pi^{\mathrm{ad}}_{[t_n, T]}$ and fixed $\th \in \Th_{\tnb}$, by monotonicity of $u(\cdot)$, we have 
\beqn 
\bextn[u(\xpiT - x^1_{t_n} + x^1_{t_n})] \leq \bextn[u(\xpiT  - x^2_{t_n} + x^2_{t_n})].
\eeqn 
Since this holds for any $\Pi^{\mathrm{ad}}_{[t_n, T]}$ and $\th \in \Th_{\tnb}$, taking first infimum over $\th \in \Th_{\tnb}$ for fixed $\Pi^{\mathrm{ad}}_{[t_n, T]}$ and then supremum over $\Pi^{\mathrm{ad}}_{[t_n, T]}$, we have 
\beqn 
V(t_n,x^1_\tn) \leq V(t_n,x^2_\tn). 
\eeqn 
Next, we show concavity of $V(t_n,x_{\tn})$. Let $0 < \alpha < 1$ and denote
\beqn
x_\tn^3 \treq \alpha x_\tn^1 + (1-\alpha)x_\tn^2\\
\eeqn
Then,  we have 
\beal
&V(t_n, x^3_\tn) \geq \sup_{\pi_1 \in \piadtn, \pi_2 \in \piadtn} \inf_{\th \in \Th_{\tnb}} \bextn\big[u(\alpha X^{\pi_1,\th_1}_T + (1-\alpha)X^{\pi_2,\th_2}_T)\big]
\eal 
Since $u$ is concave by assumption, we have 
\beal
&V(t_n, x^3_\tn) \geq \\
&\indeq \sup_{\pi^1 \in \piadtn, \pi^2 \in \piadtn} \bigg\{ \alpha \inf_{\th \in \Th_{\tnb}}\bextn \big[ u(X^{\pi_1,\th}_T)\big] \\
&\indeq + (1-\alpha)\inf_{\th \in \Th_{\tnb}}\bextn \big[ u(X^{\pi_2,\th}_T)\big] \bigg\}.
\eal 
Since the last expression is sum of two suprema, we conclude that 
\beqn
V(t_n,x^3_\tn) \geq \alpha V(t_n,x^1_\tn) + (1-\alpha)V(t_n,x^2_\tn).
\eeqn
Hence, we conclude the proof. 
\eproof

\section{Explicit Solutions with Specific Utility Functions}
We will be working with the logarithmic, power and exponential utility functions. These are of the form $\log(x)$, $x^\gam$ for $0 < \gam < 1$, $-\beta e^{-\beta x}$ with $\beta >0$ for $x > 0$, respectively, and give \textit{explicit} solutions in our robust setting. First, we give the following lemma. 
\belem
\label{lem222}
Let $v(t,x)$ be a strictly increasing and strictly concave $C^{1,2}([t_n,T] \times \mathbb{R}_{+})$ function such that $\frac{v_x}{-v_{xx}x} =c$ for some positive $c$. Then, for $\tn \leq t \leq T$, the supremum and infimum in the HJBI equation in \eqref{eqn270} are attained for
\beal
\label{eqn28100}
\mu^{*}_{t_n} &= \argmin_{\mu_t \in \Th_{\tnb}}(\lVert \mu_t - r\mathbf{1}\rVert)\\
\Sg^{*}_{t_n} &= \argmax_{\Sg_t \in \Th_{\tnb}} (\lVert \pi_t^{\intc}\Sg_t\pi_t \rVert) = C_{t_n} * I_{d\times d},
\eal
where $I_{d\times d}$ stands for the $d$-dimensional identity matrix. 
\elem
\beproof 
\beal
&\supin \big[  v_t + v_x(x\pi_t^{\intc}(\mu_t -r\mathbf{1})) + x^2\frac{1}{2}\pi_t^{\intc}\Sg_t\pi_t v_{xx}\big] \\
& = v_t + \supin \big[xv_x\pi^{\intc}_t(\mu_t - r\mathbf{1}) + x^2\frac{1}{2}\pi^{\intc}_t\Sg_t\pi_t v_{xx}\big]
\eal 
Since $v$ is $C^{1,2}([t_n,T] \times \bbR^+)$ and strictly increasing and strictly concave, the result follows by inner minimization for any fixed $\pi \in \Pi^{\mathrm{ad}}_{\tnb}$. Since $\Sg_t$ is positive definite and $v$ is concave, to minimize $\frac{1}{2}\pi_t^{\intc}\Sg_t\pi_t$, we need to choose $\Sg^*_{\tn} = C_\tn * I_{d \times d}$ for any fixed $\pi \in \piadtn$. 

Next, to find the optimal $\mu^*_{\tn}$ for any fixed $\piadtn$, we proceed as follows. We choose specifically $\tilde{\pi}_t$ as 
\beqn
\tilde{\pi}_t = \frac{1}{C_\tn}\frac{v_x(\mu^*_{t_n} - r\mathbf{1})}{-v_{xx}x}, \;\textrm{ for }\tn \leq t \leq T,
\eeqn
where $\mu^*_{t_n}$ is as in \eqref{eqn28100}. Note that $\tilde{\pi}_t$ is constant on $\tnb$, deterministic and is an element of $\piadtn$. Since $v$ is increasing with $v_x > 0$, for that $\tilde{\pi}_t$ to minimize the expression
\beqn 
v_x {\tilde{\pi}}^{\intc}_t(\mu_t - r\mathbf{1}) \;\textrm{ for }\tn \leq t \leq T,
\eeqn 
over $\mu_t$, we must choose 
\beqn
\mu^*_{\tn} = \argmin_{\mu_t \in \Th_{\tnb}}\lVert \mu_t - r\mathbf{1} \rVert, \textrm{ for }\tn \leq t \leq T.
\eeqn 
Furthermore, we have by classical minmax inequality
\beal
\label{eqn2900}
&\sup_{\pi \in \piadtn}\inf_{\th \in \Th_{\tnb}}\bigg[v_x x\pi^{\intc}_t(\mu_t - r\mathbf{1}) + \frac{1}{2}x^2\pi^{\intc}_t\Sg_t\pi_t v_{xx}\bigg] \\
&\indeq \leq \inf_{\th_{t_n} \in \Th_{\tnb}} \sup_{\pi \in \piadtn} \bigg[x v_x\pi^{\intc}_t(\mu_t - r\mathbf{1}) + \frac{1}{2}x^2\pi^{\intc}_t\Sg_t\pi_t v_{xx}\bigg].
\eal 
Next, for the right hand side of the inequality above, for a fixed $\th \in \Th_{\tnb}$ with $v_{xx} < 0$ and $\Sg_t$ being positive definite, we must have
\beqn 
\argmax_{\pi \in \piadtn}\bigg[xv_x\pi^{\intc}_t(\mu_t - r\mathbf{1}) + \frac{1}{2}x^2\pi^{\intc}_t\Sg_t\pi_t v_{xx}\bigg] = ({\Sg_t})^{-1}\frac{v_x(\mu_t - r\mathbf{1})}{-v_{xx}x}
\eeqn 
and plugging that to the right hand side of the inequality \eqref{eqn2900}, we have for $\tn \leq t \leq T$
\beal
\argmin_{\th_t \in \Th_{\tnb}}\bigg(\frac{1}{2}(\Sg_t)^{-1}\frac{v_x\lVert \mu_t - r\mathbf{1} \rVert^2}{-v_{xx}x} \bigg) = \bigg(\argmin_{\mu_t \in \Th_{\tnb}}(\lVert \mu_t - r\mathbf{1} \rVert), C_\tn * I_{d \times d}\bigg).
\eal 
But these are the values that we have plugged in and found for the left hand side of Equation \eqref{eqn2900}. Hence, again by inequality \eqref{eqn2900}, we conclude that the HJBI equation are attained for the values as in \eqref{eqn2800}. Hence, we conclude the proof. 
\eproof 

Based on Lemma \ref{lem222}, our solution methodology is, as follows. We assume first that $V(t_n,x_\tn)$ is $C^{1,2}(\tnb \times \bbR_{+})$. Then, by Lemma \ref{lem222}, we plug in the corresponding parameters for $\th \in \Th_{\tnb}$ and solve the classical Merton problem. Next, we verify that the resulting value function $V(t_n,x_\tn)$ is indeed $C^{1,2}(\tnb \times \bbR_{+})$. Hence, we will have solved the problem for $[\tn,T]$. Then, we will solve the problem for $[t_{n-1},t_n)$, and we proceed backwards up to $[0,t_1)$ via the same methodology. We emphasize here that the resulting value function $V:[0,T] \times \bbR_{+} \rar \bbR_{+}$ is not necessarily $C^{1,2}([0,T] \times \bbR_{+})$, but is a concatenation of $C^{1,2}$ functions on $[0,t_1) \times \bbR^+,[t_1,t_2)\times \bbR_+,\ldots, [t_n,T]\times \bbR_+$. 

\subsection{Logarithmic Utility Case}
First, we are going to solve the robust optimization problem with logarithmic utility $\log (x_\tn)$ and $x_\tn > 0$.
\beqn 
\label{eqn34}
V(\tn,x_\tn) = \supin \bbepn\big[\log(\xpiT)\big].
\eeqn 
We assume that $V(\tn,x_\tn)$ is $C^{1,2}(\tnb \times \bbR_{+})$ and by Lemma \ref{lem222}, we let for $t_n \leq t \leq T$
\beal
\mu^*_{\tn} &= \argmin_{\mu_t \in \Th_{\tnb}} \lVert \mu_t -r \rVert\\
\Sg^*_{\tn} &= C_\tn*I_{d \times d}
\eal 
and let 
\beqn
d X_t^{\pi} =  X_t^\pi \pi^\inc_t( (\mu^*_{\tn}-r\mathbf{1}) dt +  \sigma^*_{\tn} d W_t), \textrm { for }t_n \leq t \leq T.
\eeqn
The optimization problem reads as
\beqn
\label{eqn124}
\sup_{\pi \in \piadtn} \bextn[\log(\xpith)]. 
\eeqn
By Ito lemma, we have 
\begin{align} 
\label{eqn216}
& \sup_{\pi \in \piadtn} \bbepn\big[\log(\xpiT)\big]  \\
&\indeq 
= \log(x_{\tn}) + \sup_{\pi \in \Pi_{\mathrm{ad}}}  \bextn\bigg[ \inttnt (\pi^\inc_u(\mu^*_{\tn} - r\cdot \mathbf{1})
\\ &\indeq\indeq  -\frac{1}{2} \pisb) du \bigg].
\end{align}
Hence, by concavity on $\pi$ inside the integral, we conclude that checking first order condition inside the expectation on $\pi$ is sufficient and get that 
\begin{align}
(\mu^*_\tn - r \mathbf{1} ) - \Sigma^* \pi_t &= 0,
\end{align}
Thus,
\beal
\label{eqn129}
\pi^*_t &= (\Sigma_\tn^*)^{-1}(\mu^*_\tn - r\mathbf{1})\\
&= \frac{1}{C_\tn}(\mu^*_\tn - r\mathbf{1})
\eal
for $\tn \leq t \leq T$, and the optimal value function reads as
\beal
\label{eqn2233}
V(\tn,x_\tn) &= \bextn[\log(X^\pi_T)]\\
& = \log(x_\tn) + \frac{1}{2C_\tn}\lVert \mu_\tn^* -r\cdot\mathbf{1}\rVert^2(T-\tn).
\eal
Hence, we verify that $V(\tn,x_\tn)$ is indeed $C^{1,2}(\tnb\times \bbR^+)$ and the corresponding optimal $\piadtn$ and $\th \in \Th_{\tnb}$ for $\tn \leq t \leq T$ are
\beal
\mu^{*}_{\tn} &= \argmin_{\mu_t \in \Th_{\tnb}}(\lVert \mu_t - r\mathbf{1})\rVert)\\
\Sg^{*}_{\tn} &= \argmax_{\Sg_t \in \Th_{\tnb}} (\lVert \pi_t^{\intc}\Sg_t\pi_t \rVert) = C_{t_n} * I_{d\times d}\\
\pi^*_{\tn} &= \frac{1}{C_{\tn}}(\mu^*_t - r\mathbf{1})
\eal
Next, we go one time step backwards and examine the following optimization problem  
\beal
V(t_{n-1},x_{t_{n-1}}) &= \sup_{\piadtne}\inf_{\th \in \Th_{\tne}}\bepz_{t_{n-1}}\big[\log(X^\pi_{T})\big] \\
&= \log(x_{t_{n-1}}) + \sup_{\piadtne}\inf_{\th \in \Th_{\tne}} \bepz_{t_{n-1}} \big[ \int^T_{t_{n-1}} (\pi^\inc_u(\mu_u - r\cdot \mathbf{1}) 
\\&\indeq -\frac{1}{2}\pis du \big]\\
&= \log(x_{t_{n-1}}) + \sup_{\piadtnet}\inf_{\th \in \Th_{\tnet}} \bepz_{t_{n-1}} \big[ \int^{\tn}_{t_{n-1}} (\pi^\inc_u(\mu_u - r\cdot \mathbf{1}) 
\\&\indeq -\frac{1}{2}\pis du \big] + \supin \bextn\big[\inttnt (\pi^\inc_u(\mu_u - r\cdot \mathbf{1}) 
\\&\indeq -\frac{1}{2}\pis du \big]
\eal 
By Equation \eqref{eqn2233} for $[t_n,T]$, we have 
\beal
V(t_{n-1},x_{t_{n-1}}) &=  \sup_{\piadtne}\inf_{\th \in \Th_{\tne}}\bepz_{t_{n-1}} \big[\log(X^{\pi,\th}_{T})\big] \\ 
&= \bigg(\frac{1}{2C_{t_n}}\lVert \mu_\tn^* -r\cdot\mathbf{1} \rVert^2 \bigg)(T-\tn)\\
&\indeq + \sup_{\piade}\inf_{\th \in \Th_{[t_{n-1},t_n)}} \bepz_{t_{n-1}}\bigg[ 
\int^{t_n}_{t_{n-1}} (\pi^\inc_u(\mu_u - r\cdot \mathbf{1})\\ 
&\indeq -\frac{1}{2}\pis)du \bigg].
\eal 
Here, we apply Lemma \ref{lem222} and Theorem \ref{MOP} on the interval $[t_{n-1},t_n)$ with $t_n$ in place of $T$, $t_{n-1}$ in place of $t_n$ and $\log(X^\pi_{t_{n-1}})$ in place of $\log(X^\pi_{t_{n}})$ to the expression, 
\beqn
\log (x_{t_{n-1}}) + \sup_{\piadtnet}\inf_{\th \in \Th_{[t_{n-1},t_n)}} \bepz_{t_{n-1}}\bigg[ 
\int^{t_n}_{t_{n-1}} (\pi^\inc_u(\mu_u - r\cdot \mathbf{1}) -\frac{1}{2}\pis)du \bigg].
\eeqn
Hence, we conclude that for $t_{n-1} \leq t < t_n$
\beal
\mu^{*}_{t_{n-1}} &= \argmin_{\mu_t \in \Th_{\tnet}}(\lVert \mu_t - r\mathbf{1})\rVert)\\
\Sg^{*}_{t_{n-1}} &= \argmax_{\Sg_t \in \Th_{\tnet}} (\lVert \pi_t^{\intc}\Sg_t\pi_t \rVert) = C_{t_{n-1}} * I_{d \times d}\\
\pi^*_{t_{n-1}} &= \frac{1}{C_{t_{n-1}}}(\mu^*_{t_{n-1}} - r\mathbf{1})
\eal 
Thus, 
\beal
V(t_{n-1},x_{t_{n-1}}) &= \log(x_{t_{n-1}}) + \frac{1}{2}(\mu^*_{t_{n-1}} - r \cdot\mathbf{1})^{\intc}(\Sg^*_{t_{n-1}})^{-1}(\mu^*_{t_{n-1}} - r \cdot\mathbf{1})(\tn- t_{t_{n-1}}) \\
&\indeq + \frac{1}{2}(\mu^*_\tn - r \cdot\mathbf{1})^{\intc}(\Sg^*_\tn)^{-1}(\mu^*_{\tn} - r \cdot\mathbf{1})(T - \tn)
\eal 
Iterating backwards this way up to $[t_0,t_1)$, we have 
\beal
V(t_0, x_0) &= \log(x_0) + \sum^{n}_{i=0}\frac{1}{2C_{t_i}} \lVert \mu^*_{t_i} - r \cdot\mathbf{1}\rVert^2(t_{i+1} - t_i)
\eal 
and the corresponding optimal parameters $(\th_t)_{t_i \leq t < t_{i+1}}$ and the optimal policy $(\pi^*_t)_{t_i \leq t < t_{i+1}}$ for $t_i \leq t < t_{i+1}$ are 
\beal
\mu^{*}_{t_i} &= \argmin_{\mu_t \in \Th_{[t_i,t_{i+1})}}(\lVert \mu_t - r\mathbf{1})\rVert)\\
\Sg^{*}_{t_i} &= \argmax_{\Sg_t \in \Th_{[t_i,t_{i+1})}} (\lVert \pi_t^{\intc}\Sg_t\pi_t \rVert) = C_{t_i} * I_{n\times n}\\
\pi^*_{t_i} &= \frac{1}{C_{\ti}}(\mu^*_{t} - r\mathbf{1})
\eal 

\subsection{Power Utility Case}
We proceed to solve the robust optimization problem in power utility case. As in logarithmic utility function, following the above recipe, we assume that $V(\tn,x)$ is $C^{1,2}(\tnb \times \bbR_{+})$ and pick the corresponding $\th^{*}_\tn \in \Th_{[\tn,T]}$. 
We let for $\tn \leq t \leq T$
\beal
\mu^*_{t_n} &= \argmin_{\mu_t \in \Th_{[\tn,T]}} \lVert \mu_t -r \rVert\\
\Sg^*_{t_n} &= C_\tn*I_{d \times d}.
\eal
and solve the classical nonrobust problem
\beal
\label{eqn333}
V(\tn,x_\tn) = \sup_{\pi \in \piadtn} \bextn[ (\hxtp)^\gam],
\eal 
for $0 < \gam < 1$. The equation for \eqref{eqn333} on $[t_n,T]$ retrieved from Lemma \ref{lem222} assuming $V(t,x)$ is $C^{1,2}([0,T] \times \bbR_+)$ reads as
\beal
\label{eqn3334}
V_t + \sup_{\pi}\bigg\{ x\pi^{\intc}(\mu^*_{\tn} - r\mathbf{1})V_x + x^2\frac{1}{2C_{\tn}}\pi^{\intc}\pi V_{xx} \bigg\} &= 0\\
V(T,x) &= x^\gam
\eal
We make the Ansatz to \eqref{eqn3334} on $[t_n,T]$ for $V(\cdot,\cdot)$ along with the optimal policy  
\beal
\label{eqn3111}
V(\tn,x_\tn) &= x_\tn^\gam \exp\bigg(\frac{\gam\lVert \mu^*_\tn - r\mathbf{1} \rVert^2}{1-\gam}(T-\tn)\bigg)
\eal 
for $\tn \leq t \leq T$, which is $C^{1,2}([t_n,T] \times \bbR_+)$ and satisfies the condition in Lemma \ref{lem222} and fulfills \eqref{eqn3111}. Hence, as in logarithmic case iterating up to $t_0 = 0$, we conclude that
\beqn 
V(t_0,x_0) = x^\gam_0\bepz\bigg[\exp\bigg( \sum^n_{i=0}\frac{\gam \lVert \mu^*_{t_i} - r\mathbf{1} \rVert^2}{2(1-\gam)C_{t_i}}(t_{i+1} - t_i) \bigg) \bigg]
\eeqn for $t_i \leq t < t_{i+1}$ with $i=0,\ldots, n-1$ 
\beal
\mu^{*}_{t_i} &= \argmin_{\mu_t \in \Th_{\tib}}(\lVert \mu - r\mathbf{1})\rVert)\\
\Sg^{*}_{t_i} &= \argmax_{\Sg_t \in \Th_{\tib}} (\lVert \pi_t^{\intc}\Sg_t\pi_t \rVert) = C_{t_i} * I_{d \times d}\\
\pi^*_{t_i} &= \frac{\mu^{*}_{t_i} - r\mathbf{1}}{C_{t_i}(1-\gam)}.
\eal

\subsection{Exponential Utility Case}
We next analyze the robust utility optimization problem for the exponential utility case
\beqn
\label{eqn246}
u(x) = -\beta e^{-\beta x}
\eeqn for $x > 0$ and $ \beta >0$. 
We take $\hpi_t = \xpit \pi_t$ in \eqref{eqn15} such that
\beqn
\label{eqn249}
d\hxt = \hpi_t^\inc (\mu_t - r\cdot\mathbf{1})dt + \hpi^\intc_t\sigma_t dW_t,\; \pas
\eeqn
At $t_n < T$, the optimization problem reads as
\beqn 
\label{eqn380}
V(\tn,x_\tn) = \supin \bextn[-\beta e^{-\beta {\hxt}}]
\eeqn 
By \eqref{eqn27123}, using Theorem \ref{MOP} the HJBI equation reads as 
\beal
\label{eqnexp}
&v_{t_n} + \supin \bigg[ v_x \pi^\intc_t(\mu_t - r\mathbf{1}) + \frac{1}{2}v_{xx}\pi^\intc_t\Sg_t \pi_t \bigg] = 0\\
&v(T,x) = -\beta e^{-\beta x}
\eal
Next, we state the following result analogous to Lemma \ref{lem222}.
\belem \label{lem333}
Let $v(t,x)$ be a strictly increasing and strictly concave $C^{1,2}([\tn,T] \times \bbR_+)$ function such that $\frac{v_x}{-v_{xx}} = c$ for some positive c. Then, for $\tn \leq t \leq T$, the supremum and infimum in \eqref{eqnexp} are attained for
\beal
\label{eqn2800}
\mu^{*}_{t_n} &= \argmin_{\mu_t \in \Thtnb}(\lVert \mu_t - r\mathbf{1}\rVert)\\
\Sg^{*}_{t_n} &= \argmax_{\Sg_t \in \Thtnb} (\lVert \pi_t^{\intc}\Sg_t\pi_t \rVert) = C_{t_n} * I_{d \times d}.
\eal
\elem 
\beproof The proof is a simple modification of Lemma \ref{lem222}.  
\eproof
Based on Lemma \ref{lem222} and Theorem \ref{MOP}, we proceed to solve 
\beal 
V(\tn,x_\tn) &= \supih \bextn[-\beta e^{-\beta \hat{X}^{\hpi,\th}_T}]\\
&= -\beta \infih \bextn[e^{-\beta \hat{X}^{\hpi,\th}_T}],
\eal 
with $\hat{X}^{\hat{\pi},\hat{\th}}_T$ having the dynamics Equation \eqref{eqn249} for $\tn \leq t \leq T$
\beal
\mu^{*}_{\tn} &= \argmin_{\mu_t \in \Thtnb}(\lVert \mu_t - r\mathbf{1})\rVert)\\
\Sg^{*}_{\tn} &= \argmax_{\Sg_t \in \Thtnb} (\lVert \pi_t^{\intc}\Sg_t\pi_t \rVert) = C_{\tn} * I_{d\times d}.
\eal 
As in the previous two cases, we find $V(\tn,x_\tn)$ and verify that it is in $C^{1,2}$ as follows. Indeed,
\begin{align}
&V(\tn,x_\tn) = -\beta e^{-\beta x_\tn} \infih \bextn \big[\exp\big(\inttnt -\beta\hpi_u^\inc(\mu^*_{\tn} - r\cdot\mathbf{1})du + \intzT \hpi_u^\inc\sigma^{*}_{\tn} dW_u\big)\big]du \\
\label{eqn258}
&= -\beta e^{-\beta x_\tn} \infih \bextn \big[\exp\big( \inttnt -\beta\hpi^\inc_u(\mu^*_{\tn} - r \cdot\mathbf{1}) + \frac{1}{2}\beta^2\hpi^\inc_u \Sg^{*}_\tn\hpi_u du\big)
\end{align} 
We note that 
\beal
 -\beta\hpi^\inc_u(\mu^{*}_\tn - r \cdot\mathbf{1}) + 
 \frac{1}{2}\beta^2\hpi^\inc_u \Sg^{*}_\tn\hpi_u
\eal 
is convex in $\hpi$. Hence, by pointwise minimisation, we get that for $\tn \leq t \leq T$
\beal 
\hpi^*_{\tn} &= \frac{1}{C_\tn} \frac{1}{\beta}(\mu^*_\tn - r\mathbf{1})\\
V(\tn,x_\tn) &= -\beta e^{-\beta \xpit} \exp\bigg( -\frac{\lVert \mu^*_\tn -r\cdot\mathbf{1} \rVert^2}{2}(T-\tn)\bigg)
\eal
We see that $V(\tn,x_\tn)$ is $C^{1,2}([\tn,T] \times \bbR_+)$. Hence, the verificaton is complete. Going backwards by repeating the above verification procedure for $[t_{n-1},\tn), [t_{n-2}, t_{n-1}),\ldots [t_0,t_1)$, we conclude that the optimal parameters for $t \in [t_i,t_{i+1})$ are
\beal
\mu^{*}_{t_i} &= \argmin_{\mu_{t} \in \Th_{[t_i,t_{i+1})}}(\lVert \mu_{t} - r\mathbf{1})\rVert)\\
\Sg^{*}_{t_i} &= \argmax_{\Sg_{t} \in \Th_{[t_i,t_{i+1})}} (\lVert \pi_t^{\intc}\Sg_t\pi_t \rVert) = C_{t_i} * I_{d\times d},\\
\hpi^*_{t_i} &= \frac{1}{C_{t_i}}\frac{1}{\beta}(\mu^*_{t_i}-r\mathbf{1}),
\eal
and the value function at $(t_0,x_0)$ reads as
\beal
V(t_0,x_0) &= -\beta e^{-\beta x_0} \exp\bigg( -\sum^n_{i=0}\frac{\lVert \mu^*_{t_i} -r\cdot\mathbf{1} \rVert^2}{2C_{t_i}}(t_{i+1}- t_i)\bigg)
\eal 

\section{Concluding Remarks}
We see that the robust approach in three classical utility functions necessitates to choose the volatility of the largest magnitude with $\Sg^*_{t_i} = C_{t_i} * I_{d \times d}$ for $t_i \leq t < t_{i+1}$ for $i = 0,\ldots, n-1$ and $\Sg^*_{\tn} = C_\tn \times I_{d \times d}$ for $\tn \leq t \leq T$, whereas the drift term is to be chosen closest to the risk free interest rate with $\mu^*_{t_i} = \argmin_{\mu_{t} \in \Th_{[t_i,t_{i+1})}}\lVert \mu_{t} - r\mathbf{1}\rVert$, respectively. The optimal portion to be invested in risky assets decreases proportional to the uncertainty of $\Sg^*_t$, which is consistent with the intuition of the investor being risk-averse. A limiting argument of the uncertainty sets is also immediate by our framework. In particular, given that the uncertainty interval denoted by $\Th_t$ changes at each time $t \in [0,T]$ rather than at prespecified times $0 =t_0 < t_1 <\ldots < t_n < T$, letting the mesh $\Delta t_i \treq t_{i+1} - t_i \rar 0$, we have the optimal parameters along with the value function in exponential utility case for $0 \leq t \leq T$
\beal
\mu^{*}_t &= \argmin_{\mu \in \Th_{t}}(\lVert \mu_t - r\mathbf{1}\rVert)\\
\hpi^*_t &= \frac{1}{C_{t}}\frac{1}{\beta}(\mu^*_{t}-r\mathbf{1}) \\
\Sg^{*}_t &= \argmax_{\Sg \in \Th_{t}} (\lVert \pi_t^{\intc}\Sg\pi_t \rVert) = C_{t} * I_{d\times d}
\eal
and the value function at $(t_0,x_0)$ reads as
\beal
V(t_0,x_0) &= -\beta e^{-\beta x_0} \exp\bigg( -\int_0^T \frac{\lVert \mu^*_{t} -r\cdot\mathbf{1} \rVert^2}{2C_t}dt\bigg).
\eal 
The power and utility cases have the analogous optimal parameters and optimal values, accordingly.

\end{document}